\documentclass[twocolumn,amssymb,showpacs,aps,prl]{revtex4}
\usepackage{amsmath}
\usepackage{amssymb}
\usepackage{graphicx}% Include figure files

\newcommand{\ignore}[1]{}               %ignore enclosed text

\begin{document}

\title{Dynamical core polarization of two-active-electron systems in strong laser fields}
\author{Zengxiu Zhao}\email{zhao.zengxiu@gmail.com}
\author{Jianmin Yuan}
\address{Department of Physics,
National University of Defense Technology, Changsha 410073, P. R.
China}
\date{\today}
\begin{abstract}
The ionization of two-active-electron systems by intense laser fields is investigated theoretically.
In comparison with time-dependent Hartree-Fock  and exact two electron simulation, we show that the ionization rate is overestimated in  SAE approximation.
A modified single-active-electron model is formulated by taking into account of the dynamical core polarization. Applying the new approach  to Ca atoms,  it is found that the polarization of the core can be considered instantaneous  and  the large polarizability of the cation suppresses the ionization  by  50\% while the photoelectron cut-off energy increases slightly. The existed tunneling ionization formulation can be corrected analytically by considering core polarization. 
\end{abstract}


\maketitle

Various of  non-perturbative  phenomena occurring during atom-laser interactions  are started with single ionization, e.g.,   above threshold ionization (ATI)  and high harmonic generation (HHG). Although they have been successfully
interpreted by the rescattering model based on  single active electron (SAE) approximation (see Reviews  e.g., \cite{ Protopapas97, Brabec00}),
detailed examination showed  that multielectron effects are embedded in the photon and electron spectra \cite{Smirnova09N,Mairesse10,Spanner11,Shiner11,Boguslavskiy12,Bergues12,Pabst13}. 
It is found that high-order harmonic generation (HHG) from molecules records interference of different channels suggesting more than one molecular orbitals are involved \cite{Smirnova09N} and electron rearrangement is occuring \cite{Mairesse10},  which is certainly beyond the scope of the SAE theory.  On the other hand, two-electron events such as non-sequential  double ionization can not be explained either without considering the electron-electron interaction \cite{Fittinghoff92}. It is thus desirable to examine in details the multielectron effects occurring in the  ionization of atomic systems beyond SAE. 

The single ionization of atoms in strong laser fields can be pictured as  tunneling of one electron through the barrier formed by the atomic potential and the laser-atom dipole interaction.   The Keldysh parameter measures the ratio of tunneling time to the optical period, $\gamma=\sqrt{I_p/2U_p}$,  where $I_p=\kappa^2/2$ is the ionization potential and $U_p=E^2/4\omega^2$ is the ponderomotive energy of a free electron in a laser field of strength $E$ and frequency of $\omega$. When $\gamma<1$, tunnel ionization occurs so rapid that the electric field can be considered as a static field at each instant. The so-called adiabatic approximation is the root of Ammosov-Delone-Krainov (ADK) -like theories \cite{Ammosov86} for obtaining ionization rates. Based on this picture, the rate is mainly determined by the unitless quantity $\kappa^3/E$ with $\kappa^3$ representing the atomic field strength at the classical radius of the electron motion.

It is obvious that the adiabatic approximation will break down if the atomic potential acting on the tunneling electron is varing sooner than the tunneling time. For more than one electron systems, the core can be polarized by the laser fields, hence the atomic potential is time-varying.  In the case of absence of  resonant excitation, the polarization is instantaneously following the laser field. One therefore expects that ionization rates from single-active electron theory needed to be corrected by taking the  dynamical core-polarization (DCP)  into account \cite{Jordan08}.  Recently we have incorporated the DCP  into simulations \cite{Zhang13L} successfully interpreting the experimentally  measured alignment-dependent ionization rate of  CO molecules \cite{WuJ12}.
In this work, we  further investigate the effects of DCP on the photoelectron spectra of alkali-earth atoms that have two strongly correlated valence electrons. In particular, we benchmark the various related theories by comparison with exact solution of the time-dependent Schr\"{o}dinger equation (TDSE) for a model hydrogen molecule  with both electrons moving in one dimension.

We start with the SAE approximation and then  take into  account of the multielectron symmetry  \cite{Patchkovskii06, Santra06L,Zhao07,Zhao08b} and the core-polarization induced by laser fields.  For  a N-e$^-$ system interacting with laser  fields, the valence electrons will be strongly perturbed compared to the inner electrons. After the liberation of one electron, the ion becomes tighter bounded  giving rise to higher secondary ionization potential. Therefore  the SAE approximation is usually  adopted assuming the ionic core is frozen.  The effective  TDSE for the active electron in a laser field takes the form of 
\begin{equation}
i\frac{\partial}{\partial t}\psi=[-\frac{\nabla^2}{2}+V_n+V_L]\psi.
\label{SAETDSE}
\end{equation}
where $V_L=\vec{r}\cdot\vec{E}$ is the interaction of the active electron with the external laser field $\vec{E}$  and $V_n$ is the effective potential  from the {\it frozen} core (atomic units are used throughout unless indicated otherwise).
One of the approaches to obtain the effective potential is approximating the Hartree-Fock potential in the local density approximation, 
 that  gives the correct  asymptotic behavior of 
$V_n\rightarrow-\frac{1}{r}$ as the active electron is detached from the atomic system. The initial wave function can be  taken as the the Hartree-Fock orbital of the  valence  electron.  We will refer to this treatment as the SAE theory. 

The SAE theory assumes the electrons can be  distinguished as the  active electron  and the core electrons.
Although the static (both Coulombic and exchange) potentials from the core electrons are  taken into account,  the antisymmetrization of the total wave function due to the Pauli exclusion principle is disregarded in the dynamics driven by external fields. It can be partly remedied  by requiring the wavefunction $\psi(\vec{r},t)$ orthogonal to the occupied orbitals during the time propagation, therefore for many-electron systems, the occupied orbitals by the core electrons limit the configuration space that the active electron can occupy.   We  refer this treatment as SAE+O theory.  

Another shortcoming of SAE theory  is that it  fails when the dynamic response of the core electrons comes into play,  such as for systems that have more than one weakly bounded electrons.  The interplay between electrons would lead to complex multielectron effects including  multiorbital (multichannel) and multipole effects \cite{Pabst12}. Here we focus on 
the effect of the adiabatic evolution or polarization of the ionic core induced  by the external laser field. 
 Within the adiabatic approximation, it is possible to derive an  effective Hamiltonian of the active electron which takes into account of the laser-induced core polarization \cite{Bersuker60,Brabec05,Sukhorukov12}. We give a brief description in the follows.

Denoting the polarizability tensor of the atomic core as $\hat{\beta}^+$, the induced dipole moment is given by $\vec{d}=\hat{\beta}^+\vec{E}$, 
 where $\vec{E}$ is the external laser field.
For symmetric atomic core, the polarizability is uniform in all direction, the induced dipole moment is parallel to the external field and the potential due to laser-induced core polarization is given by \cite{Bersuker60,Sukhorukov12}
\begin{equation} 
V_{cp}=-\frac{\beta^+\vec{E}\cdot\vec{r}}{r^3}.
\end{equation}
When the active electron is close to the atomic core, the form of  polarization potential is not valid because of  the electron screening. Therefore the polarization potential is cut to zero below $r_0$ that is estimated from  the atomic polarizability ($\approx r_0^3$) \cite{Bersuker60, Zhang13L}.
It can be seen that the  magnitude of the potential  from the polarized core  is proportional to the strength of the external electric field. In strong field regime, it is comparable or larger than the interaction of the active electron with  the permanent dipole moment of the atomic core  if it exists.  

The effective TDSE for  the  active electron  turns into 
\begin{equation}
i\frac{\partial}{\partial t}\psi=[-\frac{\nabla^2}{2}+V_n+V_{cp}+V_L]\psi.
\label{TDSECP}
\end{equation}
 The method of direct  propagating Eq.~\ref{TDSECP} will be named as SAE+CP. Similar to the SAE theory discussed previously, the initial wave function is taken as the the Hartree-Fock orbital of the  valence  electron. Note in this theory,  we neglect the polarization of the core induced by the Coulombic field of the outer electron as well as the permanent dipole moment.

Different from the theories presented above, the time-dependent Hartree-Fock theory in principle takes all electrons into account based on the mean-field approximation.
We limit ourselves to the case of two valence electrons that forms a singlet state and keep the other N-2 electrons frozen.
Restricting the two electrons occupying the same orbital, and using the effective potential from the other N-2 electrons which  forms the closed-shell core, we have the following nonlinear equation,
\begin{equation}
i\frac{\partial}{\partial t}\psi=[-\frac{\nabla^2}{2}+V_{N-2}+\left<\psi|\frac{1}{r_{12}}|\psi\right>+V_L]\psi.
\label{TDRH}
\end{equation}
where $V_{N-2}$ is the effective potential from the core constituted by the other N-2 electrons, which has asymptotic behavior as $-\frac{2}{r}$. 
This method will be referred as time-dependent restricted Hartree (TDRH) method.
The repulsive Coulomb potential from the other valence electron is evaluated at each time as 
\begin{equation}
\left<\psi|\frac{1}{r_{12}}|\psi\right>=\int d\vec{r_2}|\Psi(\vec{r_2},t)|^2\frac{1}{|\vec{r}-\vec{r_2}|}
\label{V12}
\end{equation}
which includes the induced polarization from the interaction of the laser field with the other valence electron. Here we have made a crude assumption that the two valence electrons have the same time-dependent orbital.
Note that if the potential in Eq.~\ref{V12}  is evaluated with the initial field-free Hartree-Fock valence orbital, we again obtain the TDSE given in Eq.~\ref{SAETDSE}.

Now  we apply those theories to the ionization of  alkali-earth atom Ca by a laser field at wavelength of $1600$ nm, intensity of $1\times 10^{13}$W/cm$^2$. The laser pulse has a duration of 15fs with a Gaussian envelop.
The Ca atom has a configuration of $1s^22s^22p^63s^23p^64s^2$ with two valence electrons outside a closed-shell. The hartree-Fock calculation gives ionization potential of  0.1955 a.u.. The polarizability of Ca is found of 154 a.u. 
After obtaining the effective potential  from HF calculation using the local density approximation, we perform the SAE calculation and obtain the similar ionization energy  at 0.1947 a.u.  
The ponderomotive energy $U_p$ is about 3.08 times of photon energy, and the Keldysh parameter is close to 1, therefore  tunneling ionization  dominates.

The equations of motion, Eq.~(\ref{SAETDSE}, \ref{TDSECP},  and \ref{TDRH}) are solved respectively
in a spherical box of 1600 a.u. with 20 partial waves using pseudospectral grid and split-operator propagation method \cite{Tong97a,Chu00}.  The photoelectron spectra are obtained by projecting the final wavefunction to the continuum states
\begin{equation}
P_E=\sum_{l=1}^{20}\left|\left<\phi_{El}|\psi(T_f)\right>\right|^2
\end{equation}
where $\phi_{El}$ is the energy normalized continuum state of given energy ${E}$ and angular momentum ${l}$.  The single-electron ionization probability is defined as
$p_i=\int P_EdE,$
and the total ionization probability of the system is given by
$P_I=1-(1-p_i)^2,$
where independent particle approximation is applied.

\begin{figure}
\includegraphics*[width=7cm,clip=true]{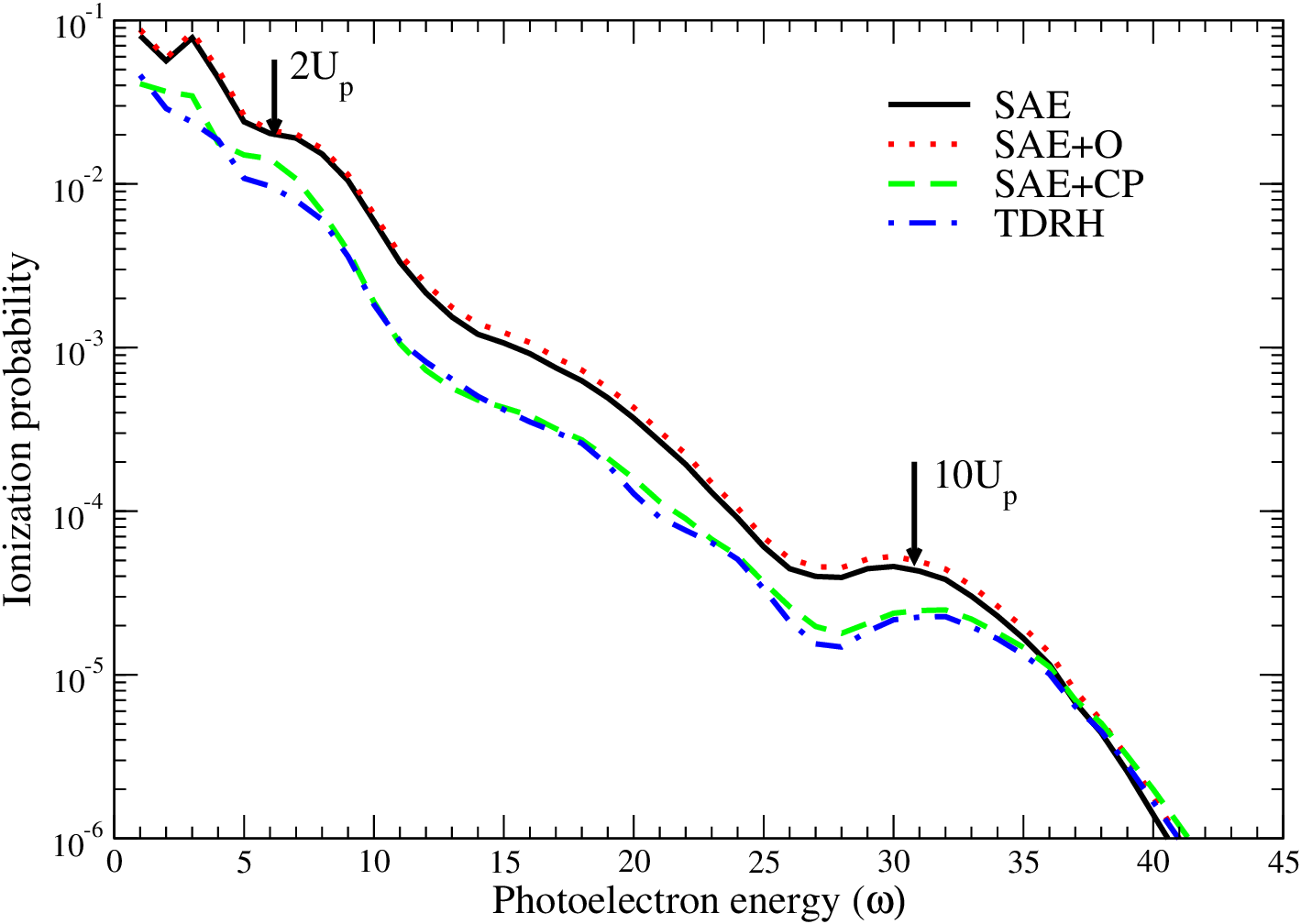}
 \caption{ Photoelectron spectra of Ca atom calculated in various approximations. The arrows indict the maximum energy of  directly escaped and rescattered electron, with U$_p$ as the ponderomotive energy.}
\label{CaPES}
\end{figure}

In Fig. ~\ref{CaPES}, we present the photoelectron energy spectra for Ca calculated with the various theories.  All calculations capture the main features of above threshold ionization  that there exists two main peaks located at $2U_p$ and $10U_p$.  The $2U_p$ peak corresponds to those electrons directly escape from the atom after tunnelling through the barrier formed by the atomic potential and the dipole interaction with the laser field.  The $10U_p$ is the maximum energy that the tunnnelling electron can gain after rescattered backward by the atomic core \cite{Paulus94}.
In the SAE+O calculation, we keep the outer electron wavefunction orthogonal to the other occupied orbitals, which implies that the core can not be penetrated. However the spectra intensity shows only a little increasing compared to the SAE calculation. 

When the core is polarized by the laser fields,  
it takes more energy for the electron to move from inside the core  to the outer region.  
However once it tunnels out the barrier, the polarization potential repels the electron away resulting slightly increase of the cut-off energy as demonstrated by the comparison of SAE results with the SAE+CP results multiplied by $2$  shown in Fig.~\ref{CaPES}. We see that  the core polarization  causes the suppression of ionization rates and  marginally  increases the maximum energy that electrons can gain.  The reason for the latter  lies in the rescattering electron dynamics \cite{Paulus94, Corkum93}. Based on the classical trajectory analysis, the laser field is close to zero when the electron collides  the atomic core with maximum kinetic energy. The corresponding instantaneous polarization of the atomic core is small thus it makes little impact on the cut-off energy of the photoelectron. On the other hand, it can be expected that the harmonic spectra  can be modified by the core-polarization due to different recombination instants and the resulted polarization as shown in \cite{Jordan08}. 

Note that we use the available polarizability of the neutral atom found in database. The polarizability of the Ca$^+$ is about $20\%$ smaller which makes small change to the ionization ratio. We also perform a different check by calculating the time-dependent induced dipole moment from propagating the Ca$^+$  in the same laser field and then use it in the simulation of the outer electron dynamics. The results show almost no difference suggesting the polarization is indeed instantaneous, therefore the phase lag due to different time response  of the outer electron and the atomic core to the laser field makes no effect in the present study. However, this might be not true when resonant excitation is present which is beyond the scope of the present study. To further justify the consideration of the core polarization, TDRH calculation is performed as well. The resulted spectra shown in Fig.~\ref{CaPES} is very close to that computed from SAE+CP calculation indicating that the core polarization can not be neglected  for the laser intensity we considered.

The suppression of ionization rate due to the core polarization can be understood within the tunneling picture.
The potential barrier along the laser polarization direction is given by
\begin{equation}
V(z)=V_{n}(z)-Ez+V_{cp}
\end{equation}
where $V_{cp}=\beta^+E/z^2$. 
In the standard theory of tunnel ionization, the core polarization is not taken into account, and the potential barrier takes the form of $V_0(z)=V_{n}(z)-Ez$.
According to the ADK  model \cite{Ammosov86}, the tunnel ionization rate when disregarding the core polarization can be estimated by
\begin{equation}
w_0\approx \exp\left[-2\int_{z0}^{z_1}\sqrt{\kappa^2+2V_0(z)}dz\right]
\end{equation}
where $\kappa$ is related to the ionization potential $I_p$ by $I_p=\kappa^/2$, and $z_0$ and $z_1$ are the inner and outer turning points respectively. When the core polarization is taken into account, the correction of the ionization rate can be approximated by \cite{Brabec05, Zhao06b, Zhao07a,Zhang10a}
\begin{equation}
R_c=w/w_0\approx  \exp\left[-2\int_{z0}^{z_1}\frac{V_{cp}}{\kappa} dz\right]
\label{Ratio}
\end{equation}
\begin{figure}
\includegraphics*[width=7cm,clip=true]{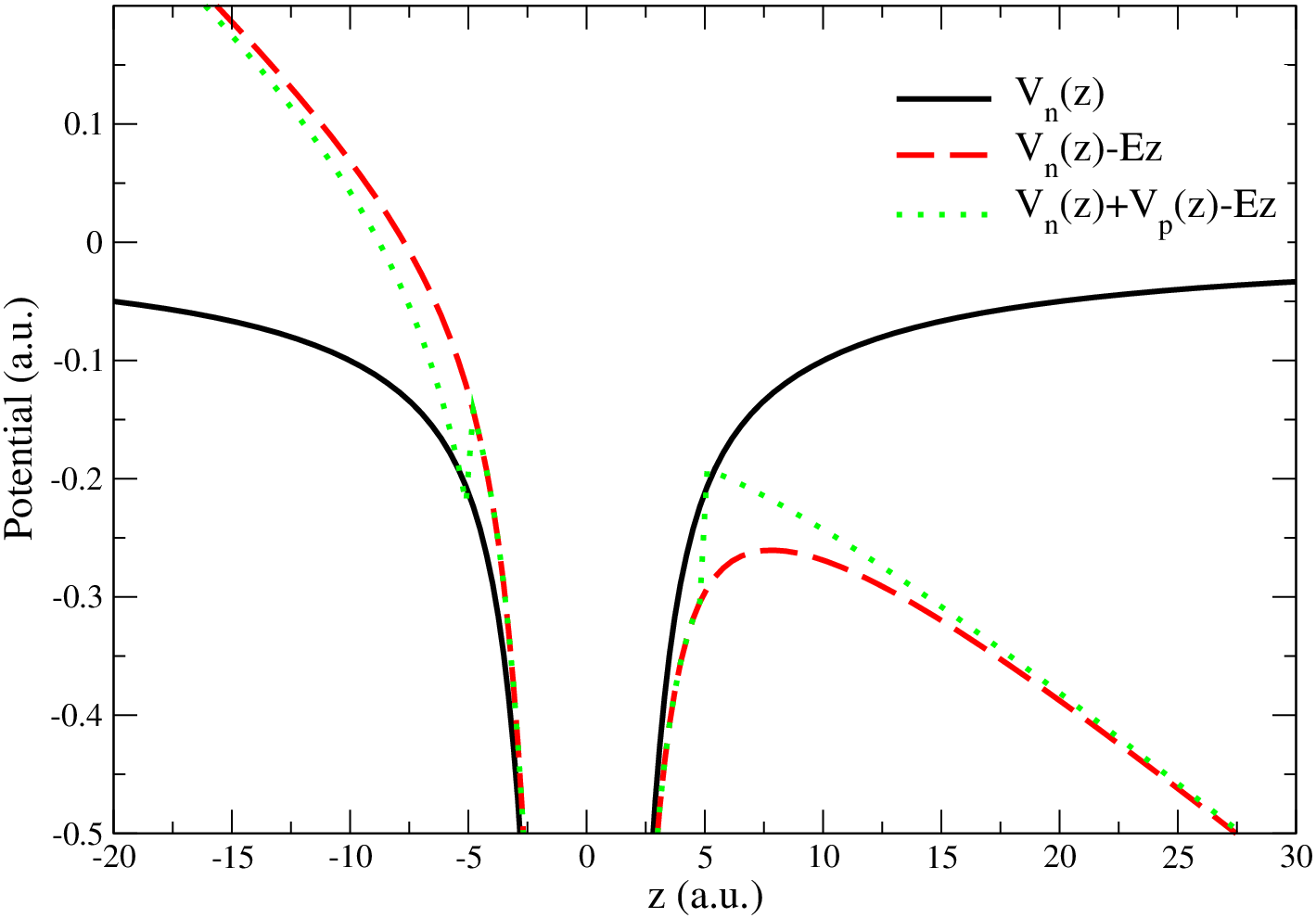}
 \caption{ Illustration of the effective potentials  for the cases of laser-free (solid line), laser field included (dashed lines) and core-polarization considered (dotted lines).}
\label{potbar}
\end{figure}

The potential barrier of Ca atom in the laser field is plotted in Fig.~\ref{potbar}.   The polarization potential is cut to zero below $r_0=5$ a.u. that is estimated from  the atomic polarizability ($\approx r_0^3$) and is consistent with the size of Ca atom.
Taking $z_0=r_0$  as the inner turning point.  The outer turning point is determined by $I_p/E$. At the  laser intensity of $ 10^{13}$W/cm$^2$, the outer turning point is 12 a.u. from the nucleus.
According to Eq.~\ref{Ratio}, we found the correction factor is  0.386, while  the numerical simulation shows that total photoelectron spectra intensity in SAE+CP  calculation is about $52\%$ of that obtained from SAE calculation in good agreement with the analytical correction.

%At laser intensity of $5\times10^{12}$W/cm$^2$ the theory predicts the correction fact as  0.44, while the numerical simulation gives 0.38. The agreement  supports the theoretical analysis. However, at the laser intensity of $2\times10^{13}$W/cm$^2$, the theory predicts  the correction factor of 0.39, while the numerical simulation gives 0.73.  At this laser intensity, the tunneling ionization model itself breaks down as the potential barrier is lower than the ground state energy.  In such regime of over-the-barrier ionization, it is known that ADK model over estimates the ionization rate by a factor of 2-4  \cite{Scrinzi99,Bauer99, Tong05}. 

\begin{figure}
\includegraphics*[width=7cm,clip=true]{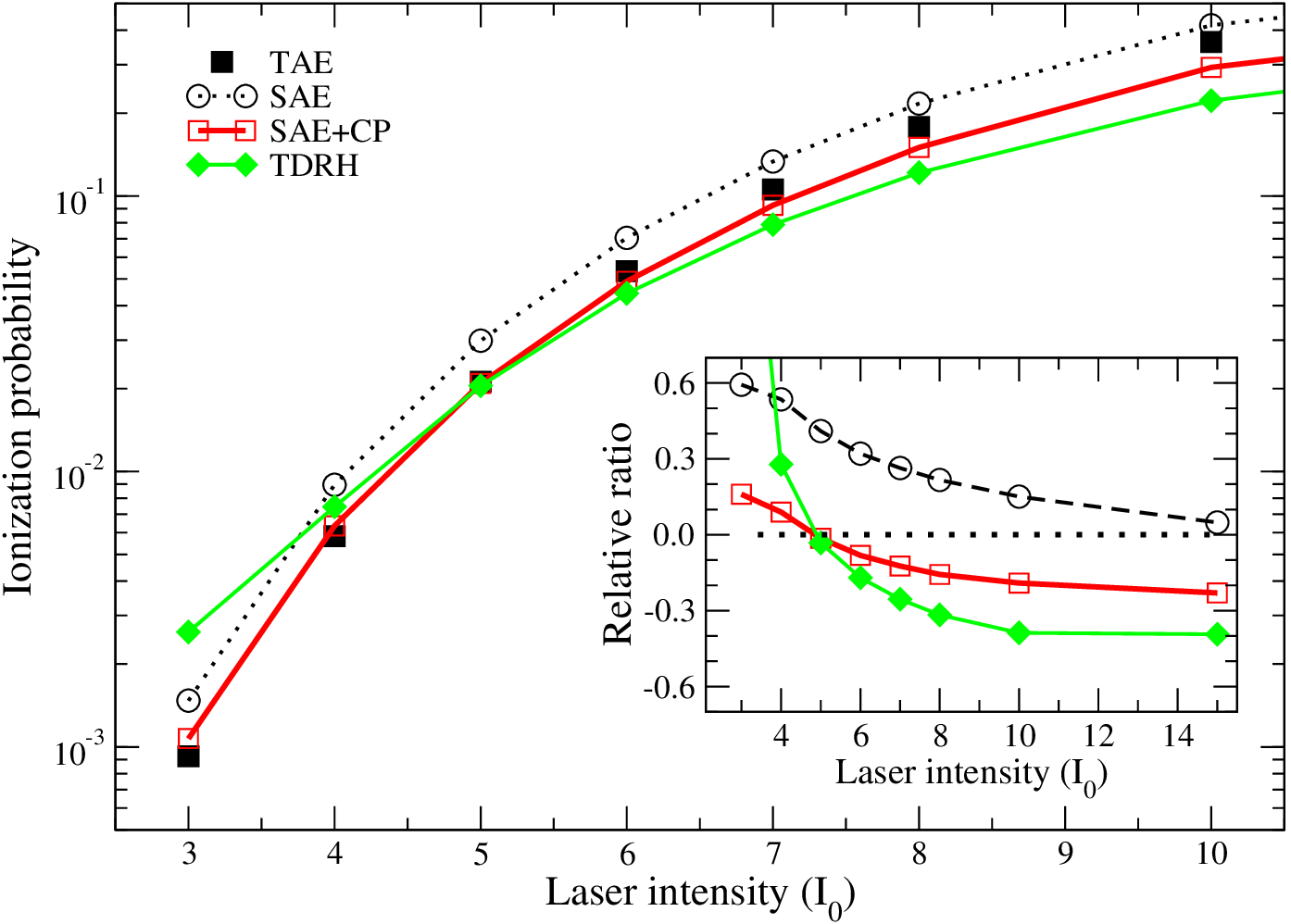}
 \caption{Ionization probability of a model Hydrogen molecule calculated from exact two-electron calculation (TAE, filled squares),   one-electron TDSE using the effective potential (SAE, dotted lines with open circles), modified SAE calculation by considering the core-polarization (SAE+CP, solid line with open squares)  and  TDRH (solid line
 with diamonds). The inset displays the relative ratios comparing to the TAE calculation from the other three theories.}
 %TDEHF (solid line with triangle) represent results from solving the time-dependent extended Hartree-Fock theory that employs two non-orthogonal electron orbitals.}
\label{SAE}
\end{figure}

In order to further check the validity of our theory,  we perform exact two-active-electron (TAE) calculation for a model hydrogen molecule with both electrons moving in one dimension. The soft-core Coulomb potential has the form of 
$|V(x)|=\frac{1}{\sqrt{\epsilon+x^2}}$ with $\epsilon_N=0.7$ and $\epsilon_e=1.2375$ for the electron-nucleus and electron-electron interaction \cite{Zhaojing10}. The energies of the neutral and the cation are found of  -2.2085 a.u. and  -1.4103 a.u. respectively. The laser pulse  has a duration of 2 cycles with wavelength of 1200 nm. The calculated ionization probability from the theories above are shown in Fig.~\ref{SAE}.
For laser intensities less than $5$$I_0$ ($I_0=10^{14}$ W/cm$^2$), the ionization probability obtained  from TDRH calculation is higher than that from  the exact  TAE calculation. The failure of TDRH at low laser intensity might be caused by two facts.  Firstly,  the HF ground state energy differs from the exact two electron energy by the correlation energy (0.025 a.u.).  When the ionization rate is  less than  the correlation energy, it has been shown that  the correlation of the two electrons has profound  effects \cite{Tolley99}. The threshold laser intensity at which the correlation can not be neglected might be estimated from that the dipole interaction energy at the classical radius  is in the order of the correlation energy. In the present studied system,   the averaged distance $\left<r\right>\approx 1.2$ a.u. such that the threshold laser intensity is roughly  $0.2I_0$.   For laser intensities much higher, the difference of  the ionization energy and the initial wave function in the restricted Hartree-Fock theory from the exact calculation can be neglected due to the strong interaction with laser filed.
Secondly,  however, the two electrons are treated as equivalent and independent particles  in TDRH which is not appropriate during ionization when the two electrons  in fact can be  distinguished as the inner and  the outer electron.  
After the removal of  the first electron, it will be very difficult to further ionize the ionic core. 
Therefore the total ionization probability defined by $P_I=1-(1-p_i)^2$ is not proper anymore and needs to be replaced by $p_i$ neglecting the ionization of second electron.

For higher laser intensity, the correlation plays less role in the ionization process. When the ionization rate is much larger than the correlation energy, the difference of the inner and outer electron orbital is not the main cause any more.  At the laser intensity of 5$I_0$, the TDRH gives the same ionization probability to the TAE calculation. However, as laser intensity keeps increasing, the rates from TDRH becomes smaller than the TAE results. The reason lies in the assumption that the two electron are  equivalent  and are occupying the same time-dependent orbital.   When ionization probability  is large, less norm  from the orbital is bounded.  The nuclear charge is less shielded and the potential on the electrons become unphysical due to this self-interaction in the TDRH theory such that further ionizing is incorrectly suppressed.  Therefore TDRH theory fails in both low and high laser intensities.

As shown in Fig~\ref{SAE}, the rates from the SAE calculation are larger than the TAE results for the laser intensity considered.  When introducing the core-polarization potential,  we see that the deviation is reduced to less than 20\%  for  laser intensity $<8$$I_0$.  Noted for laser intensity above $8$$I_0$, the ionization mechanism switches from tunnel ionization to over the barrier ionization (OTB). In the OTB regime, the removal of one electron is so rapid, there is no time for the two electrons exchanging energy such that  SAE theory works better as shown in the inset of Fig~\ref{SAE}.

In conclusion, we have shown that it is necessary to consider core polarization  in strong field ionization  of multielectron atoms. By  incorporating the polarization potential into theory the tunneling ionization theory can be improved within the frame of  SAE. Comparing with exact two-electron model calculations and time-dependent restricted Hartree-Fock calculation, we show that single-active electron approximation overestimates the ionization rate due to the additional  barrier from the dipole potential. The maximum photoelectron energy is found increased slightly by the core-polarization.

This work is supported by the National Basic Research Program of China (973 Program)
under grant no 2013CB922203,  the Major Research plan of National NSF of China (Grant No. 91121017) and the NSF of China  (Grants No. 11374366, No. 11104352). Z. X. acknowledges Dr. Xiaojun Liu for helpful discussions.

%\bibliography{/Users/Zengxiuzhao/Dropbox/zhaobibliography}

\end{document}